\documentclass[journal]{IEEEtran}
\ifCLASSINFOpdf
\else
\fi
\usepackage{algorithm}
\usepackage{algorithmic}

\usepackage{amsmath,amssymb}
\usepackage{graphicx}
\usepackage{caption}
\usepackage{subcaption}
\usepackage[keeplastbox]{flushend}
\usepackage{balance}

\usepackage{xcolor}
\newcommand{\new}[1]{\textcolor{black}{#1}}

\begin{document}
%
\title{Similarity-based prediction for channel mapping \\ and user positioning}
%
%
%

\author{Luc Le Magoarou
\thanks{Luc Le Magoarou is with bcom, Rennes, France. Contact address: \texttt{luc.lemagoarou@b-com.com}.}}

%
%

\markboth{Accepted version}{}%

%



\maketitle

\begin{abstract}
In a wireless network, gathering information at the base station about mobile users based only on uplink channel measurements is an interesting challenge. Indeed, accessing the users locations and predicting their downlink channels would be particularly useful in order to optimize the network efficiency. In this paper, a supervised machine learning approach addressing these tasks in an unified way \new{is proposed}. It relies on a labeled database that can be acquired in a simple way by the base station while operating. The proposed regression method can be seen as a computationally efficient two layers neural network \new{initialized with a non-parametric estimator}. It is illustrated on realistic channel data, both for the positioning and channel mapping tasks, achieving better results than previously proposed approaches, at a lower cost.
\end{abstract}


%
\IEEEpeerreviewmaketitle

\section{Introduction}
\IEEEPARstart{W}{ireless} networks are currently undergoing dramatic changes, driven by significant innovations in the physical layer. In particular, it has been recently proposed to use massive multiple input multiple output (massive MIMO) wireless systems \cite{Rusek2013,Larsson2014,Lu2014} with a large number of antennas in the millimeter-wave band \cite{Rappaport2013, Swindlehurst2014}, where a large bandwidth can be exploited. 

On the other hand, machine learning techniques made possible by the growing available computing power have recently led to tremendous successes in various domains \cite{Lecun2015,Goodfellow2016}. Machine learning holds promise for wireless communications (see \cite{Oshea2017,Wang2017} for exhaustive surveys). In particular, it is possible to use uplink channel data that can be acquired easily in modern MIMO systems to predict quantities of interest. In this paper, \new{the objective is to use} the channel data to predict some target vector whose content depends on the considered task. The two tasks of interest are \emph{user positioning} (in which the user's location is to be predicted) and \emph{channel mapping} (in which the user's downlink channel is to be predicted).

\noindent {\bf Contributions.} \new{This paper proposes} a similarity-based approach to tackle these two problems. \new{It is initialized with a non-parametric method but is fundamentally parametric.} Indeed, it corresponds to a neural network whose structure at the initialization mimics nearest neighbors regression \cite{Stone1977, Benedetti1977, Tukey1977}. Fine tuning by gradient descent within this structure allows to improve the prediction results. A theoretical justification of this simple method is proposed in the context of channel data. The introduced method is then empirically evaluated on realistic data and proves computationally efficient and accurate on the two considered tasks.

\noindent {\bf Related work.} User positioning and channel mapping using channel data have both been investigated. On the user positioning side, it has been proposed to tackle the problem with a convolutional neural network (CNN) operating directly on channel data \cite{Arnold2019}, or on channel data transformed to the angular domain \cite{Vieira2017}. It has also been proposed to use a classical non-parametric approach not relying on a neural network \cite{Savic2015}, operating on the received signal strengths (RSS) rather \new{than} on the channels directly. The \new{proposed method} takes advantage of the simplicity and low complexity of non-parametric approaches to guide the structure of a neural network operating on channels, which distinguishes it from all these methods.
Regarding channel mapping, which is a more recent problem, it has been tackled with help of a fully connected neural network \cite{Alrabeiah2019}. The \new{proposed method} adopts a totally different network structure, as well as a different cost function.

\section{Problem formulation}
The methods proposed in this paper apply to a wide variety of multi-user MIMO wideband systems, operating indifferently in time division duplex (TDD) or frequency division duplex (FDD), where the antennas at the base station are indifferently colocated or not (in which case it is a distributed MIMO system). Let us consider $N$ base station antennas and $S$ subcarriers, and denote $\mathbf{h} \in \mathbb{C}^{NS}$ the uplink channel vector between any given user and the base station and $h_{n,s} \in \mathbb{C}$ the channel for the $n$th antenna on the $s$th subcarrier. Note that \new{no index} is introduced to denote to which specific user corresponds the channel, since the \new{proposed method treats} indifferently the channels from all users.

Based on an incoming estimated uplink channel $\mathbf{h}$, the objective in this paper is to predict a target vector $\mathbf{t}$. Depending on the considered application, the target vector $\mathbf{t}$ may be the location of the user \cite{Arnold2019}, its downlink channel \cite{Alrabeiah2019} or any other quantity of interest. Let us denote $f$ the prediction function that maps the incoming channel to the target estimate, so that the estimated target writes
\begin{equation}
\hat{\mathbf{t}} = f(\mathbf{h}).
\label{eq:prediction}
\end{equation}
 In order to calibrate the prediction function $f$, let us assume a dataset is available containing $L$ labeled samples made of a channel and the associated target: 
\begin{equation}
\{\mathbf{h}_i; \mathbf{t}_i\}_{i=1}^L.
\label{eq:dataset}
\end{equation}
Under this very general framework, two tasks \new{are considered} in this paper.

\noindent{\bf User positioning.} The objective of this task is to predict the position of the user, based only on the knowledge of its uplink channel. In that case the target $\mathbf{t}$ contains the coordinates of the user's position (either in $2$D or $3$D depending on the context). Regarding the labeled data acquisition, it has to be obtained by an auxiliary sensor such as the GPS \cite{Savic2015}, either online or during an offline data collection phase. 

\noindent{\bf Channel mapping.} This task aims at eliminating the need for a downlink channel estimation phase. Indeed, in that case the target $\mathbf{t}$ contains the user's downlink channel, which has to be predicted from the knowledge of the uplink channel. Labeled data can be acquired during a phase where the base station sends downlink pilots, as explained in details in \cite{Alrabeiah2019}. Channel mapping is especially interesting for FDD systems that use different frequencies for the uplink and downlink transmissions. In its original formulation \cite{Alrabeiah2019}, channel mapping takes as input the channel on only a subset of the base station antennas, so as to lighten the computational cost. This setting \new{is also considered} in the experimental part of this paper.

\noindent{\bf \new{System operation.}} \new{For the two considered tasks, the system operates in two phases. During the \emph{training phase}, data are collected (using GPS or downlink pilots) and the prediction function is learned. During the \emph{exploitation phase}, the learned prediction is used. For the proposed method, these two phases are not mutually exclusive in time. Indeed, after an initial training has been carried out, they can be intertwined and online learning can be used so as to adapt to changes in the environment or users distribution.}

\section{Similarity-based prediction}

Let us now introduce the proposed prediction method, by unveiling the structure of the prediction function $f$. The method is based on the very simple rationale that similar channels lead to similar targets. The main novelty of this paper is to translate that rationale into a neural network structure instead of using a generic network structure, as was done previously \cite{Alrabeiah2019,Vieira2017,Arnold2019}.

\subsection{The Nadaraya-Watson estimator}
The similarity-based prediction \new{method proposed here} is inspired by the Nadaraya-Watson estimator \cite{Nadaraya1964,Watson1964}, \new{which is a classical non-parametric regression method}. It is based on an approximation of the joint density of inputs and targets by kernels located at the training points \cite{Rosenblatt1956,Parzen1962}. Such an approximation leads to the conditional expectation of the target given the input taking the form
\begin{equation}
\hat{\mathbf{t}} = \mathbb{E}\left[\mathbf{t}|\mathbf{h}\right] \frac{\sum_{i=1}^L K(\mathbf{h},\mathbf{h}_i)\mathbf{t}_i}{\sum_{j=1}^L K(\mathbf{h},\mathbf{h}_j)},
\label{eq:nadaraya}
\end{equation}
where $K(\cdot,\cdot)$ is a kernel function measuring the similarity between two inputs. The estimate of the target is then a convex combination of targets corresponding to the training points. In order to perform well, such a non-parametric method requires a fine enough sampling of the channel space, which may require a very large number of training points $L$. However, for the studied problem, \new{it is argued} in section~\ref{ssec:theory} that the particular nature of channel data allows to overcome this issue.

\subsection{Neural network structure}
Let us now propose a neural network structure whose initialization corresponds to the Nadaraya-Watson estimator, \new{so as to view this non-parametric method as a parametric one, whose parameters can be optimized}. To do so, a kernel measuring the similarity between channels using the complex inner product \new{is used}. Indeed, denoting $\mathcal{I}_k(\mathbf{h})$ the set of indices corresponding to the $k$ training channels that are most correlated to the current channel $\mathbf{h}$, the \new{used} kernel takes the form
\begin{equation}
K(\mathbf{h},\mathbf{h}_i) \triangleq \left\{
\begin{array}{l}
|\mathbf{h}_i^H\mathbf{h}|\, \text{ if } i\in \mathcal{I}_k(\mathbf{h}), \\
0 \,\text{otherwise}.
\end{array}\right.
\label{eq:kernel}
\end{equation}
It amounts to take into account only the $k$ training channels most correlated to the current one. The hyperparameter $k$ is chosen to be small (no more than a few dozens), yielding sparsity that reduces the computational cost of the method. With this kernel, the Nadaraya-Watson estimator can be viewed as an instance of $k$-nearest neighbors ($k$-NN) regression. 

\begin{figure}[htbp]
\includegraphics[width=\columnwidth]{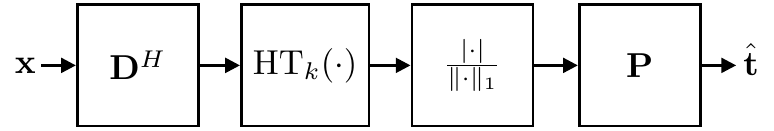}
\caption{Feedforward representation of the proposed method.}
\label{fig:net}
\end{figure}

Instead of considering this estimator as static, \new{it is possible} to view it as the forward pass in a neural network, given in algorithm~\ref{alg:sbp}, where \new{$\text{HT}_k(\cdot)$} refers to the hard thresholding operator that keeps unchanged the $k$ entries of greatest modulus of its input and sets all the others to zero. The sequence of linear and nonlinear operation can also be visualized on figure~\ref{fig:net}.

\begin{algorithm}[htb]
\caption{Similarity-based prediction (forward pass)}
\begin{algorithmic}[1] 
\REQUIRE Input $\mathbf{h}$, dictionary matrix $\mathbf{D}$,  prediction matrix $\mathbf{P}$, sparsity level $k$.
\STATE Correlation: $\mathbf{c} \leftarrow \mathbf{D}^H\mathbf{h}$
\STATE Hard thresholding: $\mathbf{s} \leftarrow \text{HT}_k(\mathbf{c})$
\STATE Normalization: $\mathbf{y} \leftarrow \frac{|\mathbf{s}|}{\left\Vert \mathbf{s} \right\Vert_1}$
\ENSURE $\hat{\mathbf{t}} \leftarrow \mathbf{Py}$ (estimated target)
\end{algorithmic}
\label{alg:sbp}
\end{algorithm}

The matrices $\mathbf{D}$ and $\mathbf{P}$ are the weights of the neural network that are to be optimized. Initializing them with
\begin{equation}
\mathbf{D} = \left( \mathbf{h}_1,\dots, \mathbf{h}_L \right),\, \mathbf{P} = \left( \mathbf{t}_1,\dots, \mathbf{t}_L \right),
\label{eq:initialization}
\end{equation}
amounts to initialize the neural network as the Nadaraya-Watson estimator.  These weights can then be optimized by gradient descent using back-propagation \cite{Rumelhart1986} to better fit the labeled dataset, according to some cost function. The precise form of the used cost functions are given in the experimental part of the paper since they depend on the considered application. \new{This fine tuning phase helps because the distribution of the training samples may not be perfect for the ultimate goal, and the used channels can be noisy. Fine tuning is meant to correct these two potential defects.}

\noindent{\bf Computational complexity.} The proposed method is particularly computationally efficient. Indeed the complexity of the forward pass is dominated by the first step (multiplication by $\mathbf{D}^H$), which costs $\mathcal{O}(NSL)$ arithmetic operations. The backward pass is even more efficient since it costs only $\mathcal{O}(NSk)$ arithmetic operations, thanks to the sparsity obtained via the hard thresholding operation. Note that this complexity is much lower than the one of concurrent approaches. Moreover, if the number of training samples $L$ is very large, it is also possible to keep only a subset of the training samples or to perform clustering to build matrices $\mathbf{D}$ and $\mathbf{P}$ in order to further reduce the complexity of the method, \new{as is classically done for radial basis function networks (RBFN) \cite{Schwenker2001}}. \new{This is one advantage of viewing the method as parametric, leading to a forward pass costing $\mathcal{O}(NS\tilde{L})$ arithmetic operations, where $\tilde{L}$ is a fixed number independent of the number of training samples.} This possibility is not explored in the current paper, \new{but an interesting avenue for future developments}. The complexity of the proposed approach is compared precisely to the one of concurrent approaches in the experimental part of the paper. 

\subsection{Why should it work?}
\label{ssec:theory}

Given the simplicity of the proposed method, it is legitimate to wonder why it should work well with channel data, although in other domains such as image or audio processing, this kind of local interpolation methods are outperformed by more elaborate deep learning methods \cite{Lecun2015}.

In order to answer this question, let us take a step back. Actually, data processing techniques are often based on the \emph{manifold assumption}: Meaningful data (signals) lie near a low dimensional manifold, although their apparent dimension is much larger \cite{Carlsson2009,Peyre2009} \cite[Section 5.11.3]{Goodfellow2016} \cite[Section 9.3]{Elad2010}. The performance of local interpolation methods is heavily dependent on the dimension of the data manifold (the lower the better). In order to perform well for a reasonable number of training points, the manifold dimension should be low. Indeed, the number of required training points for a given accuracy grows exponentially with the manifold dimension \cite{Bickel2007,Kpotufe2011}.

\emph{What is the dimension of the channel data manifold?} As previous work suggests, it is reasonable to assume the existence of a \emph{position to channel mapping} \cite{Alrabeiah2019}, i.e. a deterministic function linking the position of the user with the corresponding channel. The set of possible positions can be modeled as a two-dimensional manifold (neglecting the elevation dimension). If \new{it is} further \new{assumed} that the position to channel mapping is an homeomorphism (continuous bijection whose inverse is also continuous), then the channel vectors lie on a two-dimensional manifold (simple application of the definition of a manifold). This conclusion is very encouraging since it means that despite the high apparent dimension of \new{channel vectors} (due to the large number of antennas and subcarriers), the very low dimension of the channel manifold should allow simple local interpolation methods to perform well with a reasonable number of training points. 

Note that such a simple reasoning cannot be applied to image or audio data, and the data manifold in these domains is very likely to be much higher dimensional (e.g. a few dozens for images \cite{Lu1998}). In summary, the channel data is inherently much less complex than image or audio data, so that local methods can work well for channel data although much more computationally complex techniques have to be used in these domains (such as deep learning).

\section{Experiments}
In this section, the \new{proposed} similarity-based approach is illustrated on the channel mapping and user positioning tasks.

\noindent {\bf Implementation details.} The method is implemented with help of the PyTorch library \cite{Paszke2019}, so that gradients are computed automatically. The optimization is done with minibatch gradient descent (size of the minibatches depending on the application) using the Adam optimization algorithm \cite{Kingma2014}.
Note that complex weights and inputs are handled classically by stacking the real and imaginary parts so that the neural network treats only real numbers. \new{Moreover, for the fine tuning, the column of $\mathbf{D}$ corresponding to the current training sample is excluded in order to avoid trivial solutions during training (using for prediction a channel which is exactly the current training sample).}

\subsection{Channel mapping}
The channel mapping application is investigated with help of the DeepMIMO dataset \cite{Alkhateeb2019}, which is itself based on the ray-tracing simulator Wireless InSite by Remcom \cite{Remcom}. In order to ease comparisons with prior art, \new{the considered setting is} exactly the same as in \cite{Alrabeiah2019}, namely \new{multipath channels ($5$ paths) obtained from} an indoor environment consisting in a $10\,\text{m}\times 10\,\text{m}$ room, with $N=64$ antennas on the ceiling and $S = 16$ subcarriers, an uplink frequency of $2.4\,\text{GHz}$ and a downlink frequency of $2.5\,\text{GHz}$ (see \cite{Alrabeiah2019} for a more detailed description). The labeled dataset consists of $L$ uplink channels for which only $8$ randomly picked antennas are considered paired with the associated downlink channel (on all $64$ antennas) corresponding to users randomly located in the room, with $L$ varying in our experiment. The cost used in order to fine tune the neural network of figure~\ref{fig:net} (for $100$ epochs) is the opposite of the downlink spectral efficiency averaged over subcarriers obtained when using the estimated downlink channel as precoder, taking the form
\new{\begin{equation}
-\mathbb{E}\left[\frac{1}{K}\sum\nolimits_{k=1}^K\log_2\left( 1+ \frac{|\mathbf{h}_{D,k}^H\hat{\mathbf{h}}_{D,k}|^2}{\Vert \hat{\mathbf{h}}_{D,k} \Vert_2^2} \right)\right],
\label{eq:cost_mapping}
\end{equation}}
where $\mathbf{h}_{D,k}$ is the downlink channel on the $k$th subcarrier, $\hat{\mathbf{h}}_{D,k}$ is its estimation given by the neural network, and the expectation being in practice estimated by averaging over minibatches of $1000$ channels. The hard thresholding parameter is fixed to $k=5$. 

\begin{figure}[tbp]
    \centering
    \includegraphics[width=\columnwidth]{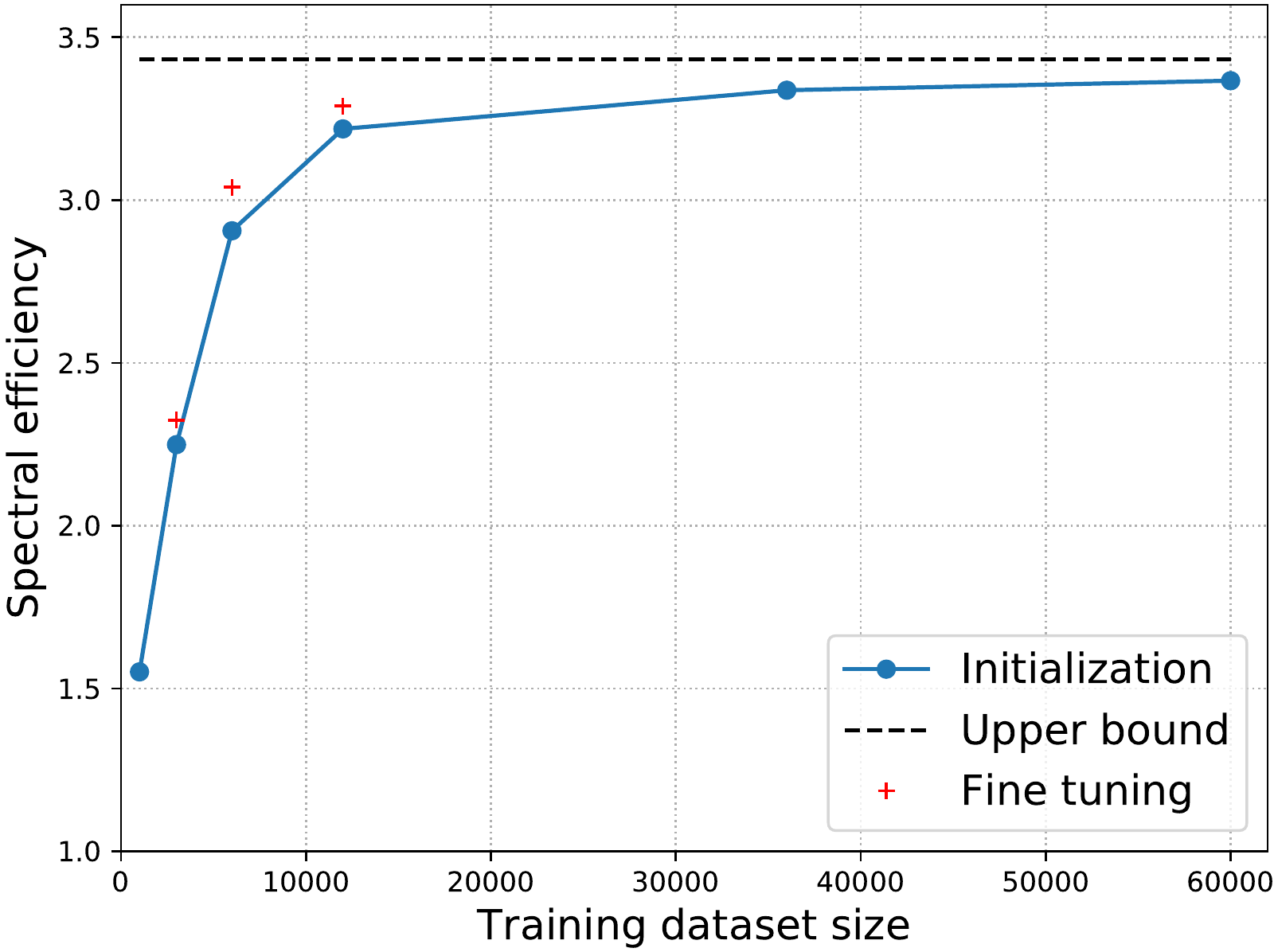}    
    \caption{Channel mapping results.}
    \label{fig:channel_mapping} 
\end{figure}

\noindent {\bf Results.} Figure~\ref{fig:channel_mapping} summarizes the obtained results for $L \in \{1000, 3000, 6000, 12000, 30000, 60000\}$, showing the spectral efficiency (opposite of the cost~\eqref{eq:cost_mapping}) averaged over $1000$ randomly picked test channels. Results are shown for the initialization (blue dots) and after fine tuning by gradient descent for $L \in \{3000, 6000, 12000\}$ (red crosses). An upper bound (obtained using the true channel as precoder) is also shown (black dashed line). It is interesting to notice that only a reasonable amount of training data is required to get very close to the upper bound with the proposed method. Indeed, the obtained results are within $6\%$ of the upper bound without fine tuning and within $4\%$ with fine tuning, with only $12000$ training points. This performance is superior to the previously proposed method \new{\cite[Figure~7]{Alrabeiah2019}}, which obtained results within $9\%$ of the upper bound with $120000$ training samples \new{(ten times more)}. Moreover, the proposed method yields a much lighter network (forward pass costing around $1\mathrm{e}6$ arithmetic operations) compared to the fully connected neural network of \cite{Alrabeiah2019} (around \new{$3\mathrm{e}7$} arithmetic operations). In summary, the proposed method yields better results with fewer training samples, at a lower cost. This experiment shows the great potential of the similarity-based approach for the channel mapping problem. \new{However, the number of required training samples may still seem large in regard of the quite small area considered in this experiment. This issue should be taken into account in future work. Note that this issue is less significant in the next experiment regarding the user positioning task.}

\subsection{User positioning}
 The user positioning task is investigated with help of a dataset originally intended for the IEEE Communication Theory Workshop (CTW) 2020 data competition \cite{CTW}, built with a channel sounder described in \cite{Arnold2019}. The considered base station comprises $N=56$ antennas and communicates over $K=924$ subcarriers at the central frequency of $1.27\,\text{GHz}$. The labeled dataset consists of $L = 4979$ channels $\mathbf{h}$ with the corresponding location $\mathbf{p}$ that are measured in an area of approximately $1\,\text{km}^2$ in the streets of Stuttgart, Germany. It is divided into $4096$ channels used for training and $883$ channels used as validation data.

\noindent {\bf Dimensionality reduction.} Channels described in the previous paragraph are of very high dimension. Indeed, they can be seen as $56\times 924$ complex matrices. Using directly the method of algorithm~\ref{alg:sbp} on these channels would result in a very high computational complexity. In order to reduce the computational burden, the dimensionality of input channels \new{is reduced} by computing the left singular vector corresponding to the largest singular value and using it as input for algorithm~\ref{alg:sbp}. This results in inputs of size $N=56$ \new{complex numbers ($112$ real numbers)} only. The cost used to fine tune the network of figure~\ref{fig:net} (for $50$ epochs) corresponds to the average localization error, taking the form
\new{\begin{equation}
\mathbb{E}\left[\left\Vert \mathbf{p} - \hat{\mathbf{p}} \right\Vert_2\right]
\end{equation}}
where $\mathbf{p}$ is the true position and $\hat{\mathbf{p}}$ its estimation given by the neural network, and the expectation being in practice estimated by averaging over minibatches of $100$ channels. The hard thresholding parameter $k$ is varied between $2$ and $16$.

\begin{figure}[tbp]
    \centering
    \includegraphics[width=\columnwidth]{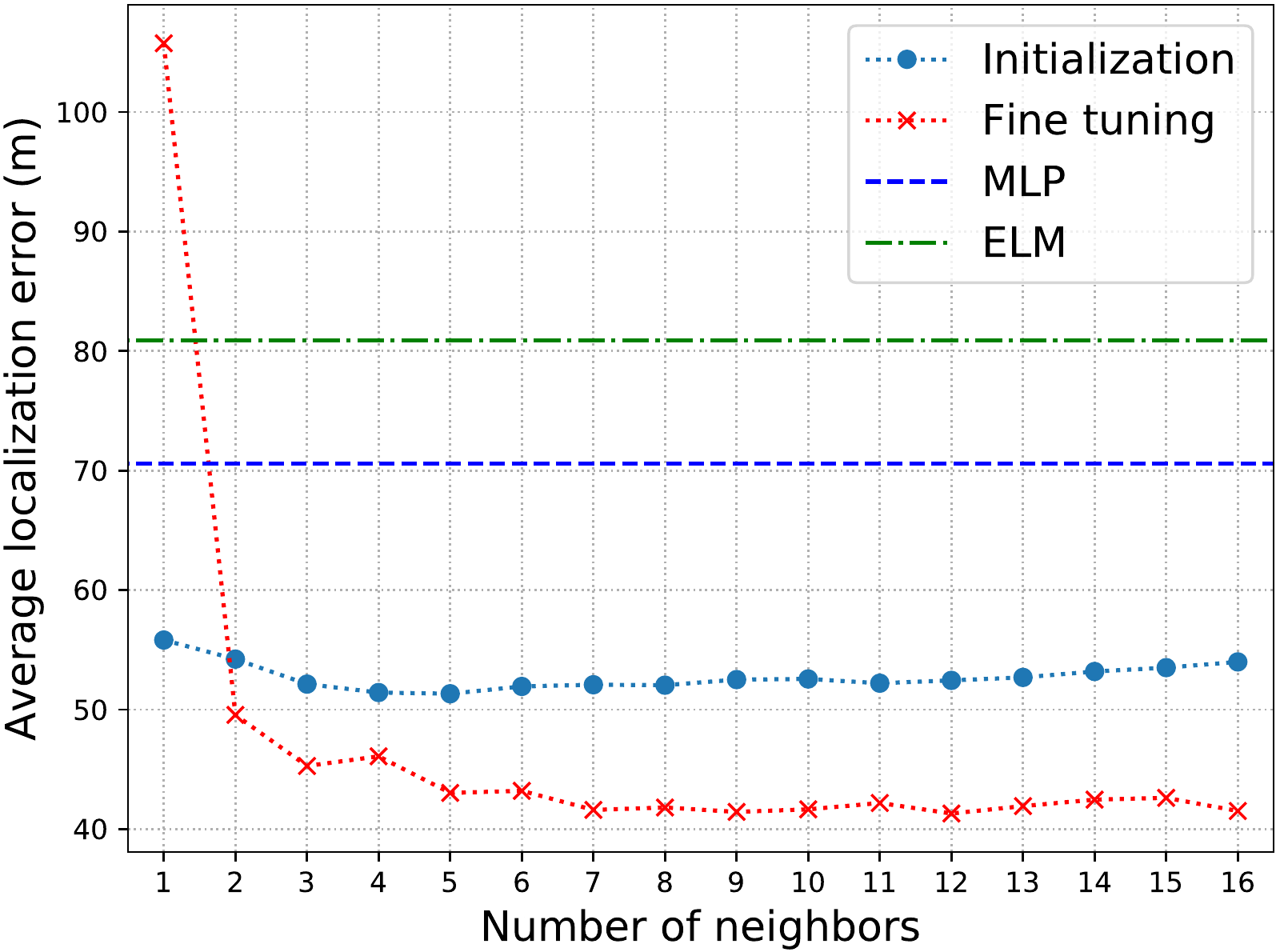}    
    \caption{\new{User positioning results.}}
    \label{fig:localization} 
\end{figure}

\noindent {\bf \new{Baselines.}} The proposed method is compared to two baselines taking the same inputs. The first one is an instance of extreme learning machine (ELM) as was proposed for user positioning \cite{decurninge2018}, with a random layer of $50000$ neurons and a rectified linear activation (ReLu) \cite[Chapter 6]{Goodfellow2016}. The second one is a simple multilayer perceptron (MLP) \cite[Chapter 6]{Goodfellow2016} comprising two layers made of $112$ neurons with a ReLu non-linearity followed by a last linear one with three neurons (outputting location estimates). It is trained for $200$ epochs with the Adam optimization algorithm \cite{Kingma2014}.

\noindent {\bf Results.} Figure~\ref{fig:localization} summarizes the obtained results for a varying number of neighbors (corresponding to the parameter $k$ of the hard thresholding operator) \new{and for the two baselines}. First of all, \new{note that the proposed method leads to better positioning results than the baselines (the MLP leading to approximately $70\,\text{m}$ and the ELM to approximately $80\,\text{m}$ average errors). Then,} it is interesting to notice that the number of neighbors has only a modest influence on the localization error, provided it is chosen not too small (larger than $4$). Second, fine tuning shows beneficial for the localization task \new{(except for $k=1$, where the fine tuning decreases a lot the positioning accuracy)}. Indeed, the initialization gives errors around $52\,\text{m}$ while fine tuning allows to attain errors around $42\,\text{m}$ (approximately $20\%$ better). Keeping in mind that the area on which localization is sought is of $1\,\text{km}^2$, such results are pretty accurate and encouraging. Moreover, \new{note that} most of the time the localization error is much smaller than the average (the median is around $20\,\text{m}$ \new{for $k=10$}). Note that these results are the first to be reported for this outdoor localization task. Concurrent parametric methods such as convolutional neural networks were only applied to indoor localization for now \cite{Arnold2019}.


\section{Conclusion}
In this paper, a generic similarity-based neural network was introduced in order to operate on channel data, for the channel mapping and user positioning tasks. It was theoretically motivated relying on the manifold hypothesis. Moreover, the proposed method was empirically validated on realistic data for the two aforementioned tasks.
In the future, it would be very interesting to investigate further the sample complexity of the method, in order to optimize the training of such systems. Moreover, one can also envision using several heuristics aimed at improving the training of such similarity-based neural networks \cite{Schwenker2001}. 

 \balance

\ifCLASSOPTIONcaptionsoff
  \newpage
\fi



\bibliographystyle{IEEEtran}
\bibliography{biblio}
\end{document}